# Securing the Objectivity of Relative Facts in the Quantum World
Richard Healey

## 1. Introduction

This paper compares and contrasts relational quantum mechanics (RQM) with a pragmatist view of quantum theory [1–3] that I'll call *desert pragmatism* (DP). DP was first developed in the Sonoran desert of Tucson: and by denying that quantum theory itself introduces any new physical objects or magnitudes (Bell's "beables") it satisfies the philosopher Quine's taste for desert landscapes. I'll first explain important points of agreement. Then I'll point to two problems faced by RQM and sketch DP's solutions to analogous problems. Since both RQM and DP have taken the Born rule to require relative facts I next say what these might be. This brings me to my main objection to RQM as originally conceived—that its ontology of relative facts is incompatible with scientific objectivity and undercuts the evidential base of quantum theory. In contrast DP's relative facts have all the objectivity we need to accept quantum theory as scientific knowledge. But a very recent modification to RQM [21] has successfully addressed my main objection, bringing the two views into even closer alignment.

## 2. Where we Agree

The function of quantum states marks a major dividing line between different ways of understanding quantum theory. On one side of the line there are those who take the quantum state to play an important role in saying what the physical world is like. They typically hold that a single, privileged quantum state describes the entire universe: other quantum states may be assigned to its subsystems (or branches), but these all ultimately derive from this one universal quantum state. Among other things, such a universal state may be taken to be a physical field in some high-dimensional configuration space, a way of specifying the dynamics describing particles or classical fields, or a *sui generis* entity on which all else supervenes.

  On the other side of the line are those who deny that there can be a universal quantum state to play any such descriptive role, because a quantum state has the very different function of specifying probabilities for different possible events involving a system, contingent on its interaction with some *other* system or systems (including in a measurement on the system). Their rivals may object that quantum cosmology would be impossible if there were no universal quantum state. But the objection fails since no realistic quantum cosmology could hope to assign a quantum state to a system with the vast number of degrees of freedom we find in the universe, so any feasible quantum cosmology would apply only to a subsystem with a much smaller set of degrees of freedom, capable of affecting a distinct "internal" subsystem.

  Along with QBism and other views, RQM and DP both regard the primary function of a quantum state to be that of specifying probabilities for different possible events involving a system and something else. So it is potentially misleading to say that a quantum state assignment specifies what state a system is in, as if a quantum state were an intrinsic, non-relational property of that system.

  At this point, an important difference emerges, between QBism on the one hand and RQM and DP on the other. For QBists, both quantum state assignments and Born probabilities are personal: each agent using quantum theory assigns their[1] own coherent

---

[1] I use the term 'their' not just to be gender-neutral, but because a QBist agent need not be a single human—it may be a group of people, or anything else capable of decision and action based on relevant information about the world, such as a robot equipped with a

degrees of belief to the possible outcomes of their measurements. They take a quantum state to be relational because it relates the system to which the state is assigned to the agent assigning it. It follows that there would be no quantum states if there were no agents capable of using quantum theory. A QBist sees no need to define or model an agent or a measurement physically. In her view, *agent* and *measurement* are treated as primitive notions within quantum theory, which she takes to be an empirically-motivated extension of personalist Bayesian probability theory. But an agent might also apply that theory to another agent, by modeling that agent as a physical system to which to assign a quantum state.

The concept of an agent plays no fundamental role in relational quantum mechanics, according to which the quantum state of a target system is simply a relation to a second physical system of any kind with which it has interacted. The target system has that quantum state (relative to the second system) whether or not there are any agents capable of assigning it. But if there happens to be an agent able to know a system's quantum state, this agent may use this knowledge to calculate the probabilities of various possible physical events involving that system, including outcomes of measurements on it.

DP sides with RQM in denying that the concept of an agent plays any fundamental role in quantum theory. A quantum state is not relative to an agent. Systems have (relational) quantum states in worlds without agents, they had them in this world before there were any agents, and they would still have them if all intelligent beings were suddenly wiped out from the universe. Just as in RQM, the quantum state of a system is relative to something physical. But this other *relatum* is not a physical system, but a specific kind of physical situation.

This appears to mark a significant divergence from RQM. But in RQM a system has a state relative to a second system only when these systems have interacted: without this interaction, it has no relative state. So for RQM the second *relatum* is not simply another system: it is the situation of that system after the two systems have interacted. RQM characterizes this situation functionally when Rovelli uses the word 'information' in saying what it is (according to a third system $P$) for a system $O$ to have information about another system $S$:

> From the point of view of the $P$-description, the fact that the pointer variable in $O$ has information about $S$ (has measured $q$) is expressed by the existence of a correlation between the $q$ variable of $S$ and the pointer variable of $O$. The existence of this correlation is a measurable property of the $S – O$ state.
> ([4, p.1652], [5, p.9])

DP also functionally characterizes the kind of situation relative to which a system has a quantum state by calling this an agent-situation. Any agent that happened to be in this situation would be in a position to physically access information sufficient to assign a quantum state to a system relative to that situation. But an agent-situation is not an agent, and it may be described in purely physical language whether or not any agent is actually in this situation. The situation is described in part by specifying a bounded space-time region that may be thought of as where some localized agent might be for some period while accessing the information and assigning the state: this restricts the accessible information to the past light-cone of that region. But there may be further physical restrictions on accessible information due to the physical isolation of that space-time location from sources of information necessary to assign this state. In the paradox of Wigner's friend [6], for example,

---

sufficiently sophisticated artificial intelligence.

Wigner is physically isolated from his friend while he remains outside the friend's laboratory.

DP and RQM each claim to offer realist views of quantum theory despite their joint denial that a quantum state is, or constrains, an element of physical reality. What they take to be physically real is not the quantum state, but properties of quantum systems. Some of the statements to which the Born rule assigns probabilities turn out to be true while others are false. The statements are assertions about the values of magnitudes—that the system has a certain energy or spin-component, for example. Many are true when no agent performs a measurement, and few of them concern macroscopic systems or their observable features. The true ones state what RQM calls "facts". The views are realist because these statements are about the physical, mind-independent world, not about what any observer or agent experiences or does. In section 4 I will explain an important disagreement between DP and RQM on the nature of these statements. But both views agree that they describe physical reality, and that it is the role of quantum states to specify, for each of a set of incompatible statements about a physical system, the probability that it is true.

Despite their realism about physical properties and the physical systems that have them, neither view fits the following (overly narrow) condition tailored to conform to the ontic models framework of Harrigan and Spekkens [7]:

(Single-world) **Realism**: The system has some physical properties, a specification of which is called its *ontic state*, denoted $\lambda$. Ontic states take values in a (measurable) set called the *ontic state space.* [8, p.6]

DP and RQM do not meet this condition. They agree that the only physical properties in quantum theory are those to which the Born rule assigns probabilities, and that these are not specified by an ontic state that takes values in any set with a non-trivial probability measure defined on it. But both views are realist in a broader, philosophically relevant sense.

**3. Two Problems in RQM**

According to RQM as originally conceived, quantum states and facts are relative in the same way: but according to DP, while quantum states are relative, facts are not relative in the same way. I will say in section 5 what I mean by 'fact', and 'relative fact'. Here I'll review the way the notions of 'state', 'fact' and 'relative fact' have been used in RQM and raise two problems for RQM before explaining solutions offered by DP in section 4.

While Rovelli's initial presentation of RQM ([4], [5]) does not use the terms 'relative fact', and 'relative state', the terms are used freely in recent restatements of the view ([9], [10], [11]). This has helped to clarify RQM without significantly changing it. So I'll first quote key early passages and then show how these terms have recently been used to clarify it (all page numbers are from [4]).

> Quantum states, as well as values of physical quantities, make sense only when referred to a physical system (which I denote as the observer system, or reference system). (1650)

> By using the word "observer" I do not make any reference to conscious, animate, or computing, or in any other manner special, systems. (1641)

> For a fixed observer, the eigenstate-eigenvalue link is maintained. (1673)

> ...a quantum mechanical description of a certain system (state and/or values of physical quantities) cannot be taken as an "absolute" (observer-independent) description of reality ...Quantum mechanics can therefore be viewed as a theory about

> the states of systems and values of physical quantities relative to other systems. A quantum description of the state of a system *S* exists only if some system *O* (considered as an observer) is actually "describing" *S*, or, more precisely, has interacted with *S*. The quantum state of a system is always a state of that system with respect to a certain other system. More precisely: when we say that a physical quantity takes the value *v*, we should always (explicitly or implicitly) qualify this statement as: the physical quantity takes the value *v* with respect to the so and so observer. (1648)

> ...in quantum mechanics, "state" as well as "value of a variable"—or "outcome of a measurement"—are relational notions in the same sense in which velocity is relational in classical mechanics. We say "the object *S* has velocity *v*" meaning "with respect to a reference object *O*." Similarly, I maintain that "the system is in such a quantum state" or "$q = 1$" are always to be understood "with respect to the reference *O*." In quantum mechanics all physical variables are relational, as velocity is. (1649)

While two notions of 'state' figure in RQM, each has a distinctive function. One is the quantum state, often denoted by a state vector |ψ>. For RQM, this is simply an instrument for calculating transition amplitudes for processes and probabilities for events that happen at interactions: a quantum state does not denote an independent element of physical reality. But there is another notion of state that consists in a quantum system's having dynamical properties, such as a particular value of a dynamical variable. As the quotes illustrate, presentations of RQM are sometimes ambiguous between these two notions of state. But the third quote may be thought to offer a defense of this ambiguity by linking the two notions.

Proponents of RQM have insisted that both the quantum state of a system and an event of one of its dynamical variables taking on a value are relative to some other system with which it has interacted. In more recent terminology, these are now called respectively a *relative state* and a *relative fact*.

"In RQM, facts determine states, not the other way around." [10, p.2]. Relative facts were important for RQM because until very recently they were taken to form the basic ontology of quantum theory. This is a narrowly-tailored notion of fact. RQM takes it to be facts in a broader sense (assuming non-relativistic quantum mechanics) that electrons exist and have spin ½ , that $qp - pq = i\hbar$, and that the hydrogen atom Hamiltonian takes the particular form it does: these latter facts are not relative but absolute in RQM. Di Biagio and Rovelli [10] assume that Hamiltonians are not relative in their response to the first problem for RQM, raised by Pienaar [13, pp.10–12].

The problem arises when quantum systems *F*, *S* interact in such a way that their final entangled joint vector state (relative to quantum system *W*) is expressible as a biorthogonal decomposition in either of two incompatible basis pairs. Di Biagio and Rovelli [10] object to Pienaar's use of the term 'measurement' in describing this case, preferring the more neutral term 'interaction'. But in their terminology, the problem is to say why this interaction gives rise to facts (relative to *F*) about eigenvalues of eigenvectors in one basis pair rather than another. In response, Di Biagio and Rovelli appeal to the dynamics of the *F*, *S* interaction (its Hamiltonian) to specify to which (relative) facts it gives rise. But if the quantum state of *F – S* (relative to *W*) just predicts *W*'s probabilities, then it has no dynamics given by a non-relative physical Hamiltonian. So RQM cannot use the Hamiltonian to describe the dynamics of an interaction that gives rise to relative facts about some observables but not others. As noted in [14], there are even cases where the interaction Hamiltonian has a structure permitting no discrimination between facts about eigenvalues of non-commuting observables because it is expressible in terms of operators transforming between vectors from either of their respective bases.[2]

This problem points to a more basic difficulty. Presentations of RQM claim that relative facts occur consequent upon *any* interaction, irrespective of its dynamics:

> For relative facts, every interaction can be seen as a "Copenhagen measurement", but only for the systems involved. Any physical system can play the role of the "Copenhagen observer". [11]

But what could these relative facts be? If they assign a precise value to every observable on a system then RQM risks conflict with non-contextuality no-go theorems such as [15]. But to take the selection of observable to be random would be to introduce a novel and otherwise unmotivated stochastic element into quantum mechanics. Despite the claim to universality, it seems that only a class of special interactions select the relative facts to which they give rise, and this class is never precisely specified in RQM.

RQM faces a second problem: at what *time* during an interaction of the right kind does the event corresponding to a relative fact occur? Physically realistic interactions gradually diminish but never completely cease. Rovelli [4, p.1652] responded to

> the well-known and formidable problem of defining the "precise moment" in which the measurement is performed, or the precise "amount of correlation" needed for a measurement to be established—see for instance Bacciagaluppi and Hemmo (1995). Such questions are not classical questions, but quantum mechanical questions, because whether or not $O$ has measured $S$ is not an absolute property of the $O-S$ state, but a quantum property of the quantum $O-S$ system, that can be investigated by $P$, and whose yes/no answers are, in general, determined only probabilistically. In other words: *imperfect correlation does not imply no measurement performed, but only a smaller than 1 probability that the measurement has been completed.* (Italics in the original)

But this response fails to address the problem, once one acknowledges that an entangled state of the $O$-$S$ system can play no role in describing its sub-systems' properties, and so cannot answer the question as to whether $O$ has completed a measurement on $S$ by recording its outcome as a fact relative to $O$ about the value of the "pointer" observable. $P$'s investigation by a measurement on the $O-S$ system would inform him of facts relative only to the $P$ system, with no clear relevance to this question. Rovelli s ([12]) further defense of this operational criterion fails to solve the problem for the same reason.

More recent elaborations of RQM do not address these problems. Di Biagio and Rovelli [9, p.30] simply say

> In the early history of quantum theory it was recognised that every measurement involves an interaction, and it was said that variables take values only upon measurement.
> RQM notices that every interaction is in a sense a measurement, in that it results in the value of a variable to become a fact. These facts are not absolute, they belong to a context. And there is no 'special context': any system can be a context for any other system.

And in [10, p.7] they say

> Q: When does something become a fact? A: Something becomes a fact, relative to you, when you interact with a system.

If a quantum state does not describe any physical property of a system, then its evolution can play no role in specifying what kind of relative facts arise in a system's interactions or the moment or period during an interaction when an event becomes a fact relative to that system.

---

[2] I thank a referee for drawing this preprint to my attention.

**4. Two Pragmatist Solutions**
According to RQM as originally conceived, relative facts concern values taken on by magnitudes at the conclusion of an interaction between two quantum systems. But RQM then faces the problems of clearly specifying which type of interaction results in which relative facts, and specifying just when this happens. These are problems even if facts are absolute.

In addressing analogous problems, DP appeals both to quantum models of decoherence and to a view of meaning called pragmatist inferentialism [1, 2].[3] The problems are analogous, not identical: intuitively, there are fewer relative facts according to DP because most interactions don't give rise to relative facts. There is a relative fact only in the perspective provided by a situation involving extensive environmental decoherence. This restriction is not arbitrary: it follows from the requirement that a statement about the value of a magnitude have enough content to be assessed for truth from that perspective.

According to pragmatist inferentialism [16], a statement derives its content from its inferential relations to other statements, and ultimately to perception and action. Important content-conferring inferences are generally reliable but not deductively valid: they are what Sellars [17] called material inferences. In an interference experiment like that of [18] a statement about the precise position of a molecule in the interferometer supports few reliable inferences connecting it to other statements and so has little content: therefore it should not be assessed for truth or falsity. By contrast, a statement about the position of a molecule on their silicon detection screen supports many reliable inferences and so has high content. Environmental position-decoherence of its quantum state at the screen marks a context relative to which it is appropriate to say that some statement quite precisely locating a molecule on the screen is true.[4]

By assigning a quantum state this additional function DP conflicts with what Di Biagio and Rovelli ([10, p.1]) explicitly say:

> In RQM, the quantum state is not a representation of reality: it is always a relative state and is only a mathematical tool used to predict probabilities of events relative to a given system.

But DP and RQM agree on the main point: a quantum state does not represent an element of physical reality and is not what Bell called a beable.

In DP a statement about the value of a property on a quantum system is meaningful only when that system's interaction with its environment may be modeled in quantum theory as involving robust and stable environmental decoherence of the system's reduced quantum state with respect to a "pointer basis" of eigenstates of the operator corresponding to that property. In such a context a statement that the system has the property supports many reliable material inferences and so has a rich enough content to be assessable for truth or falsity. This model of decoherence is used not to *represent* the fact that the system has or that it lacks that property, but only to license one to treat the statement that it has the property as either true or false in that context and to try to find out which by making suitable observations. Juffman *et. al* [18] observed the positions of their molecules on the detection screen by imaging them in a scanning-tunneling electron microscope.

---

[3] By contrast, Di Biagio and Rovelli [9] appeal to decoherence only to explain how what they call *stable* facts arise in the quantum world. For them, a stable fact arises to the extent that the probability of a measurement outcome may be expressed as a sum of probabilities, each calculated by conditioning on a different possible outcome of a hypothetical intermediate measurement of an incompatible observable. This is essentially equivalent to neglecting interference terms in calculating the overall probability. A stable fact need not be a relative fact.

[4] Chapter 12 of my [2] shows how pragmatist inferentialism offers an analog rather than digital model of content, defending it against objections and explaining how it helps us to understand the meaning of statements that arise in quantum mechanics.

   The first problem for RQM was to specify the type of interaction required for a particular kind of relative fact to result. DP solves its analogous problem by pairing a resulting relative fact about a property of a system with a subspace spanned by elements of the "pointer basis" of the model of decoherence that makes a statement about that property meaningful and possibly true. Although environmental decoherence is extraordinarily rapid it is never perfect: the pointer basis continues to fluctuate slightly. So the property corresponding to a subspace also fluctuates a little. This is fine, because the basis is not used to represent exactly which relative fact obtains but only to license application of the concepts of truth and falsity to statements about properties of a system in a context. How well these concepts apply is a matter of degree just as is the content of statements to which they might be applied. So a model of decoherence is not required to single out *precisely* those properties that may be assessed for truth or falsity in a context: the range of relative facts may safely remain slightly indeterminate.

   The second problem for RQM was to answer the question of exactly when a relative fact results during an interaction of the right type. DP solves its analogous problem by rejecting a presupposition of the question. There is no moment during an interaction modeled by environmental decoherence after which the pointer basis ceases to fluctuate and remains permanently fixed. Decoherence is never permanent, even though it may be extremely long-lasting. But again this does not matter, because a magnitude's having a property on a system as a result of an interaction does not require that property to be paired with exactly one fixed subspace of the system's Hilbert space.

**5. What are Relative Facts?**

   Whatever we mean by a relative fact, that notion must make sense whether or not quantum theory is true. So we can arrive at an understanding by seeing how to apply the notion first in a classical world.

   As I type these words I am in Spain two feet from a window. Here are two sentences I now use to state these facts:

   α. I am in Spain.     β. A window is two feet from me.

Which of these is a relative fact? I shall argue that neither fact is relative.

   The first statement is true only because at the time the sentence was typed the person typing it was in Spain. If I were to utter the same sentence while in Canada I would not state a fact, because that statement would be false. Anyone else asserting it in Spain would then be stating a different fact. What fact (if any) a person typing, writing, speaking or otherwise asserting α states depends on the context in which it is asserted. This context depends on when, where and by whom the statement is made. But when this context has been specified, the statement's content is fixed, and if it is true it states a fact that is not relative to anything else.

   Like the first sentence, the second sentence is relational. When I typed sentence β I stated a fact, because at the time the relation denoted by the expression "two feet from" held between the person typing it and the window. If I were to move away from the window and type the same sentence again I would state no fact because that relation would no longer hold. Alternatively, I could use β to make a false statement by moving the window further away from me. It does not follow from the relational character of sentence β that any fact I state by typing β is a relative fact. The sentence "Two is greater than one" is relational, but anyone asserting it would be stating the same fact in any context—a fact that is not relative to anything else.

There is a different reason for thinking that any fact stated by a true assertion of β is a relative fact. Spatial distance implicitly depends on a specification of reference frame. In the case of β, the (approximately inertial) frame in which the surface of the earth is at rest provides a tacitly assumed reference frame. In relativity (unlike Newtonian physics) the distance between me and my window is relative to a reference frame.

But is that reason good enough? Sentence α contains the indexical term 'I' that functions to refer to the person using α to make a statement by asserting it. The context of assertion determines that it refers to the one who asserts it—absolutely and not relative to anything else. An indexical term like 'now' has the analogous function of referring to when a person uses a sentence containing that term. Although no such temporal indexical appears explicitly in either α or β, they are both in the present tense. In its tensed use, the use of the verb 'to be' may be taken implicitly to supply the temporal context of an assertion of one of these sentences, and thereby explain why I make different statements, one true and the other false, when I assert either of them now in Spain and later in an open field in Canada. Moreover, in relativity the context in which I typed β may be taken to determine that this use of the present tense implicitly selects the moment (or interval) of time *in my frame* as I typed the sentence. So there is really nothing relative about the facts I stated by asserting them both now. Each assertion is true, and the fact it states is absolute and not relative to anything else.

The theory of relativity shows other ways in which a specification of reference frame determines which fact is stated by a true assertion. It shows that what statement is made by asserting a sentence specifying the time order of a pair of spacelike separated events, or the strength of a magnetic field, also varies with reference frame. Except in unusual circumstances, the context of assertion fixes the frame, thereby determining the content of the statement and so what non-relative fact (if any) it states.[5]

For example, it is said that each superconducting magnet in the main storage ring at CERN exerts a magnetic field of magnitude 7.7 Tesla.[6] The context that determines the content of this statement, and so the fact it states, includes a reference frame that the magnetic field strength is implicitly referred to—an (approximately inertial) reference frame in which the magnets are at rest. Referred to the instantaneous rest frame of a passing proton, a different fact may be stated by saying that the magnet exerts a magnetic field strength of some 53,000 Tesla.

There is no tension between these two facts because in their rest frame the protons also experience a strong counteracting electric field. Neither is a relative fact, because the term 'magnetic field strength' refers to different magnitudes when used in these two different contexts (as does 'electric field strength'). But each fact may be stated by asserting the appropriate sentence in the context of either frame. For example, in the context of the proton frame, the fact about the 7.7 Tesla magnetic field may be stated by saying that the magnetic field strength is 6930×7.7 Tesla and the electric field strength is 6930×7.7×$v$, where $v$ is the proton velocity relative to the magnet (just 3.1 m/sec less than the velocity of light). The theory of relativity does not introduce any relative facts. It merely requires one to recognize new ways in which a reference frame explicitly or implicitly affects how non-relative facts are stated by the assertion of sentences whose content is partly determined by the context of assertion.

---

[5] In his [19] Fine touts a contrary view in which the content of such a statement requires no specification of frame because it is already "tensed". This is not the appropriate place to dispute Fine's fragmentalist metaphysics of special relativity. I merely note that it does not involve relative facts and has no natural extension to general relativity.

[6] This is when protons circulate in the ring with energy 6.5 Tev.

We have seen that neither the relational form of a sentence nor the phenomenon of linguistic contextuality that affects what statement is made by asserting such a sentence gives rise to anything deserving the name of a relative fact: nor does the advent of the theory of relativity. If we are to find relative facts in the quantum world we must look elsewhere, since all the facts we have encountered so far are not relative to anything beyond what is stated by an assertion of a sentence in a context—to what I called the content of that assertion. But this assumed that whether such an assertion states a fact depends only on its content and not on the context in which the statement is *assessed* as true rather than false. If the truth-value of a statement made by asserting a sentence in context $c_1$ can vary as a function of the context $c_2$ in (or respect to) which its truth-value is assessed, then what is stated by an assertion assessed as true in $c_2$ may be called a relative fact if in a different context of assessment $c_2{*}$ that same statement would not be assessed as true.

MacFarlane ([20]) calls this phenomenon assessment sensitivity. He argues that it occurs in several areas of language use, including disagreements based on personal taste. Yum, for example, loves to eat liquorice, while Yuk hates the taste and so never eats liquorice. Suppose Yum utters the sentence γ. Liquorice is tasty.

Yum thereby makes a statement that he takes to be true—to state a fact. But Yuk disagrees: he takes Yum's statement to be false. According to MacFarlane, they agree on the *content* of Yum's statement (what Yum asserted in this context), but disagree because Yum assesses it as true while Yuk assesses it as false. Tastes may change: but at the time of their dispute, Yum's taste provides a context of assessment in which his assertion of γ is true, while Yuk's taste provides a different context of assessment in which Yum's assertion of γ is false. So it is a fact that liquorice is tasty relative to Yum but not relative to Yuk: this is what makes it a relative fact.

MacFarlane's treatment of this and other examples is controversial among philosophers of language. But it does at least give us something that was lacking until now—an analysis of what it could mean to talk of relative facts. MacFarlane begins his 2014 book by calling it a defense of a coherent form of relativism about truth. Since truth and facts are two sides of the same coin, it may also be considered a defense of a coherent notion of a relative fact. The idea behind this notion is that an assertion states a relative fact if and only if it is never simply true, but merely true when assessed in a certain context. If MacFarlane is right, then in asserting sentence γ, Yum stated a relative fact because what he said was true when assessed according to his standard of taste but false when assessed according to Yuk's.

From now on I'll call truth as assessed in a context of assessment *relative truth*. I'll say that a statement is *perspectival* just in case it has a truth-value (is true or is false) only relative to an appropriate context of assessment. A perspectival statement states a *perspectival fact* in a context of assessment if and only if it is relatively true in that context. I'll use the term *relative fact* to denote a perspectival statement that is relatively true in a context of assessment; and I'll call a perspectival statement that is relatively true in one context of assessment but false in another a *strongly relative fact*.

## 6. RQM Relative Facts are not Strongly Relative

How well does this notion of a relative fact square with what more recent formulations of RQM say about relative facts in quantum theory? Recall the following quote [9, p.30]

> RQM notices that every interaction is in a sense a measurement, in that it results in the value of a variable to become a fact. These facts are not absolute, they belong to a context. And there is no 'special context': any system can be a context for any other system.

Elsewhere in [10] they adopt Pienaar's [13] term 'perspective' in characterizing RQM and use it to refer to the set of all relative facts for a system in a context. Since they use the terms 'fact' and 'event' interchangeably, a perspective may equally be identified with the set of all relative events for a system in a context. By an event, RQM means a dynamical variable taking on a value on a physical system, or a system having a dynamical property. The Born rule assigns probabilities to these events, and in a measurement the outcome records the occurrence of an event (though an event may occur in an interaction that is not literally a measurement, because the interaction was neither arranged nor its outcome recorded by any agent or observer).

If we take the term 'context' in RQM to refer to a context of assessment, then RQM may be understood to make the key claim that all facts are relative according to the analysis of section 5.[7] RQM takes any statement about an event to be perspectival when it is made by asserting a sentence Σ ascribing a value $q$ to a dynamical variable on a system. RQM claims that such a statement expresses a relative fact in section 5's sense of 'relative fact'. But it is not a strongly relative fact because it turns out that, according to RQM, when such a statement is true in one context there is never another context in which it is false.

In an example related to the paradox of Wigner's friend [6], Di Biagio and Rovelli [10, p.5] suppose that a "friend" $F$ and a qubit $Q$ are physically linked through a "measurement interaction" in the computational basis while $W$ initially interacts with neither $F$ nor $Q$. I understand their initial presentation of this scenario to involve interactions of a particular kind among generic quantum systems $Q$, $F$, $W$ and not just measurements by human observers. The outcome *1* of $F$'s interaction with $Q$ is then a relative fact in $F$'s context $c_F$, but the statement *St* that $F$'s interaction has outcome *1* can be assessed neither as false nor as true in a context $c_W$ in which $W$ has interacted neither with $Q$ nor with $F$. When $W$ does subsequently interact similarly with $Q$ and with $F$, that constitutes a new context $c_W*$ relative to which (ideally) there will be two new relative facts—the outcome of $W$'s interaction with $Q$ and the outcome of $W$'s interaction with $F$. The entangled quantum state of the system $Q - F$ relative to $W$ predicts that these two outcomes will be correlated. If the outcome of $W$'s interaction with $Q$ is *1* then this is a relative fact in $c_W*$. But it is not the fact stated by *St*, both because it concerns a different, later event and because it is relativized to a different context.

Assume instead that while the outcome *1* of $F$'s interaction is a relative fact in context $c_F$, in the context $c_W*$ of $W$'s subsequent interaction with $Q$ the outcome is *0*, an outcome that the quantum state of the system $Q - F$ relative to $W$ predicted to be equally likely. Then it is the *0* outcome of $W$'s interaction with $Q$ that is a relative fact in $c_W*$. Suppose $W$ were to infer that the outcome of *F*'s interaction was *0*, so the statement *St* is true in $c_F$ but false in $c_W*$. This would be a mistake in RQM according to [10] since quantum theory does not license that inference: *W* cannot even infer the outcome of *F*'s interaction from the outcome of *W*'s interaction with *F*. The outcome of *F*'s interaction is a fact only in *F*'s perspective, while the outcome of *W*'s interaction on *F* is a fact only in *W*'s perspective.

---

[7] As a referee noted, some statements of RQM contain passages that suggest alternative readings of 'relative fact' including those rejected in the previous section. But recent restatements ([10], [11]) emphasize the radical difference from more familiar kinds of relationality that "RQM does not preserve the idea that consistency can be established between different observers' accounts" [10, p.10].

The conclusion, that if *St* states a relative fact in one context then there is no context in which not-*St* states a relative fact, depends on a general feature of RQM. In RQM as originally conceived, inferences from one fact to another are legitimate only within a perspective. Because this feature is general, the argument generalizes to apply to all applications of RQM. RQM relative facts are not strongly relative.

**7. Relativism**

The main problem for RQM as originally conceived is that its ontology of relative facts involved a form of relativism that is incompatible with scientific objectivity. I'll introduce this problem by pressing an objection of Pienaar [13], explaining why Di Biagio and Rovelli's ([10]) response is inadequate. Using section 5's analysis of the notion of a relative fact, I'll show why RQM involved a variety of truth-relativism that conflicts with the basic requirement that scientific facts be objective. This problem has now been successfully addressed in a very recent preprint [21] which proposes a substantial modification to RQM. Finally I'll explain how DP is able to secure objectivity in quantum theory even in the face of recent arguments based on extensions of the *Gedankenexperiment* of Wigner's friend.

Pienaar [13] objects to RQM on the grounds that it is committed to an ontology of what he calls "island universes" (i.e. perspectives). Di Biagio and Rovelli accept Pienaar's account of RQM's ontology but object to this name he gives it because it falsely suggests that different perspectives cannot be compared. This is what Di Biagio and Rovelli ([10, p.7]) say about the ontology of RQM:

> [Pienaar] correctly characterises RQM's view: there are facts relative to every system, but that the different perspectives on reality, namely, the ensemble of facts relative to a single system, cannot be compared in an absolute manner; they can only be compared via a physical interaction. This is correct.

In explaining how physical interaction permits comparison they continue by giving a modified example (based more closely on Wigner's own scenario) in which the qubit has been replaced by a generic quantum system *S*.

> Consider on the case in which the systems *F* and *W* are actually "observers" in the rich sense of the term. Say they are humans with laboratories, notebooks and books that store and process knowledge about the world. Let us focus on *F*. What is the meaning of the statement that *F* has knowledge about the world, for instance about *S*? There are two possible answers. The first is a naturalistic answer. The second is a dualistic or idealistic answer. According to the first, this is a statement about the actual physical configuration of the ink and the notebooks, the charges in the computers and the synapses in the brain in *F* and about the correlation of these with whatever can be observed in *S*. According to the second, *F*'s knowledge is something over and above its physical configuration. In this case, the "inaccessibility" of *F*'s knowledge, namely of the "universe as seen by *F*" is indeed there. But this only follows because one assumes that knowledge is unphysical. We adhere to a naturalistic philosophy. In a naturalistic philosophy, what *F* "knows" regards physical variables in *F*. And this is accessible to *W*. If knowledge is physical, it is accessible by other systems via physical interactions. It is precisely for this reason that knowledge is also subjected to the constraints and the physical accidents due to quantum theory. Here naturalism is opposed to dualism and idealism. This conception of naturalism is closely related to what Price ([22], [23]) calls the ontological doctrine of object naturalism:

> As an ontological doctrine, it is the view that in some important sense, all there *is* is the world studied by science. [23, pp.4–5])

The ontological object naturalist, like Di Biagio and Rovelli, is committed to the view that if there is such a thing as knowledge, it is ultimately something physical. But they take naturalism to involve the even stronger commitment that the physical embodiment of knowledge is accessible via physical interactions. Unfortunately for Di Biagio and Rovelli, their own examples show that this additional commitment cannot be met.[8]

Can *W* access the physical embodiment of *F*'s knowledge about *S* in their modified example? That physical embodiment consists of the physical configuration of the ink and the notebooks, the charges in the computers and the synapses in the brain in *F* and about the correlation of these with whatever can be observed in *S*. These are all perspectival facts in *F*'s perspective, although it would be extremely hard for *F* to access the synapses in her own brain. To access any of these perspectival facts, *W* must measure a variable on *F*, *S* or something else in *F*'s hitherto isolated laboratory. Given the quantum state of this entire laboratory relative to *W*, the Born rule predicts the probability of each possible outcome of *W*'s measurement (relative to *W*). If *W* has made a joint measurement of multiple variables on different subsystems of the laboratory, the Born rule yields probability 1 that their values will embody in *W*'s laboratory concordant "records" of the outcome of *F*'s measurement on *S*. These "records" now constitute perspectival facts in *W*'s perspective. But they are not records embodying *W*'s knowledge of the outcome relative to *F*.

Section 6's discussion of Di Biagio and Rovelli's original Wigner's friend example makes it easy to see why. If *S* is a qubit *Q* then *W's* predicted probability is ½ for either possible outcome of his measurement on any subsystem of *F*'s laboratory that embodies *F*'s knowledge of her own outcome. So *W*'s relative outcome is completely uncorrelated to *F*'s relative outcome of her measurement on *Q*. Since a record must be correlated with what it records, none of *W*'s "records" is a record embodying knowledge of *F*'s relative outcome. So by interacting with anything in *F*'s laboratory *W* can acquire no knowledge of *F*'s outcome. Even *F*'s sincere answer to *W*'s question as to what that outcome was leaves *W* completely ignorant about *F*'s outcome. Pienaar's term 'island universe' is an apt metaphor for this situation. *F* and *W* are each confined to their own epistemic islands since their perspectives contain no common perspectival facts. Each of them may come to believe they share common knowledge as a result of mutual interactions, but this is an illusion. Neither can access the perspectival facts of the other. All either can do is to use their interactions to tell a consistent story of what is contained in the other's perspective that may bear no relation to its actual contents.

---

[8]In a 2018 conversation Rovelli objected that DP's talk of agents puts it in conflict with naturalism. As section 2 explained, according to DP a quantum state is not relative to an agent but to an agent-situation. Because an agent-situation may be described in purely physical terms, DP appears to meet Di Biagio and Rovelli's condition of naturalism. It is true that the description in section 2 was not purely in terms of quantum systems and facts about them. It talked about physical isolation and other physical barriers to informational access without saying how these are embodied in non-perspectival quantum facts. And it appealed to the light-cone structure of a relativistic space-time that may not be describable within a fundamental theory of quantum gravity. But this is still consistent with Di Biagio and Rovelli's strengthened form of ontological object naturalism. Only if their notion of naturalism were strengthened still further to require a physical description of an agent-situation purely within a fundamental quantum theory would DP fail to count as naturalist. But that would put all of current science in conflict with naturalism, a consequence I consider a *reductio ad absurdum* of this super-strong naturalism.

In its original form, RQM's ontology consisted of physical systems and certain relative facts about them. A relative fact concerning system $F$ results from an interaction between $F$ and another system $S$. It is a perspectival statement that is then true in $F$'s context but not in that of another system $W$ that has not interacted with $S$ or $F$. Truth in such a context is merely relative. RQM's original ontology of relative facts commited it to a variety of truth-relativism. This may be a coherent view, but it is not compatible with scientific objectivity.

This is now effectively acknowledged in the very recent preprint [21] proposing a substantial modification to RQM with the implication that "the set of 'quantum events' should be regarded as absolute, observer-independent features of reality in RQM, although quantum states remain purely relational" [21, p.8]. This brings RQM into closer accord with DP. The proposal is to modify RQM by permitting meaningful absolute matching of descriptions of events relative to different systems in accordance with a postulate of

> *Cross-perspective links.*
> In a scenario where some observer Alice measures a variable *V* of a system S, then provided that Alice does not undergo any interactions which destroy the information about *V* stored in Alice's physical variables, if Bob subsequently measures the physical variable representing Alice's information about the variable *V*, then Bob's measurement result will match Alice's measurement result. [21, p.5]

This new postulate makes it possible for $W$ to access $F$'s outcome of her measurement on $Q$ by performing his own measurement, either on $Q$ or on $W$, with the assurance that (in the absence of intervening information-destroying interactions) $W$ and $F$ will agree on the outcome of $F$'s measurement. A proponent of the modified form of RQM can now hope that the absolute outcomes of similar quantum measurements can provide objective data supporting belief in quantum theory. But recent arguments based on scenarios extending that of Wigner's friend provide an independent reason for thinking that the outcome of a quantum measurement is merely a relative fact [24]. So we still need to know what could be meant by talk of relative facts and how such talk can be squared with the objectivity of science.

Notions of truth-relativism and objectivity can be applied much more widely than just to views of quantum mechanics since they concern truth and facts in general. Plain non-relative notions of truth and fact may be characterized by two principles.

*Truth*          A statement that *P* is true if and only if *P*.
*Fact*           A statement states a fact if and only if it is true.

Corresponding relative notions may be characterized by these alternative principles.

*Relative Truth* A statement that *P* is true-relative to-*c* if and only if *P*-relative to-*c*.
*Relative Fact*   A statement states a fact-relative to-*c* if and only if it is true-relative to-*c*.

Here *c* is a context in which a statement is assessable, not a context in which it is made. A variety of truth-relativism is associated with a class of statements and contexts for which no plain notion of truth and fact is applicable, but only notions of relative truth and relative fact.

It follows from the plain concepts of truth and fact that a statement that *P* states a fact if and only if it is true; that is, if and only if *P*. In accordance with these concepts, a fact is a fact without regard to perspective. This irrelevance of perspective makes it natural (if overblown) to call a plain fact absolute or *transcendently objective*: it is absolute insofar as it is not relative to anything like a context or viewpoint, and transcendently objective as it is not limited by such things. In contrast, a relative fact in perspective $P_c$ is what is stated by a perspectival statement that is true in $P_c$—true relative to the context $c$ of $P_c$. In general, what is true relative to $c$ may not be true relative to $c^*$: it may even be false relative to $c^*$, and so strongly relative. But a relative fact in perspective $P_c$ may be a relative fact in the perspective $P_{c*}$ of *every* context $c^*$ in which it is assessable. That would not make this a plain fact because it is not simply true: the plain notion of truth is not applicable to a perspectival statement. It would rather make it a statement of (what may be called) an *immanently objective fact*. Science depends on immanently objective facts because these are what all sincere inquirers can come to agree on no matter what context they are assessed in. If there were no immanently objective facts then there would be nothing scientists would be justified in acknowledging as *data*, capable of confirming or refuting scientific knowledge claims. Science depends on immanent objectivity but not transcendent objectivity.

    It was because the relative facts in RQM's original ontology are not immanently objective that RQM's original truth-relativism is incompatible with scientific objectivity. Prior to introduction of cross-perspective links, RQM adopted a variety of truth-relativism according to which a statement about a dynamical property of a physical system has a truth-value only relative to the context of another physical system after the two systems have suitably interacted. The second system may have the right kind of organized complexity to constitute a scientist or other observer. If it does, the interaction may record a relative fact about the first system for that observer system. But it does not record a fact relative to any other observer system. A subsequent interaction with another observer system may record a further relative fact about the first system for another observer system. But that further relative fact has nothing to do with the original observer's recorded relative fact. Conceptually, there is no way even to make sense of the idea that both observers have recorded the same dynamical property of the first system, even if these records are communicated among observer systems by whatever physical interactions are involved in normal means of communication such as speech and writing. So adopting this version of truth-relativism prevented RQM's relative facts from being the publicly accessible observation reports that scientific objectivity requires of data if they are to support scientific knowledge, including quantum theory. If the original version of RQM were right then no data would be capable of providing the objective evidence for quantum theory.[9]

    By contrast, DP [25] is able to secure the objectivity of scientific knowledge by adopting a less radical variety of truth-relativism that relativizes truth of a statement about a system's dynamical property to a more restricted set of special contexts associated with environmental decoherence. An *M-decoherence environment E* of a quantum event $e$ in which magnitude $M$ may be said to take on a value in a system is a region of spacetime $R_E$ that includes the region where $e$ occurs, together with physical processes in $R_E$ that can be modeled by robust decoherence of that system's states in a "pointer basis" associated with

---

    [9]The notion of a *stable* fact may be thought to provide an answer to this objection [9, pp.2–4]). Environmental decoherence of the right kind may justify speaking of a stable fact on system $F$ for system $W$ even when this is not a fact relative to $W$. But decoherence does not select *which* property of $F$ is the stable fact from a partition of possible dynamical properties of $F$. If $W$'s subsequent interaction with $F$ makes one of these properties a relative fact for $W$ in that later context, quantum theory gives $W$ no reason to retrospectively identify this with one rather than another prior stable fact for $W$. So the decoherent emergence of a stable fact does not render this accessible to any physical system except $F$, let alone to a complex physical system with the right structure to constitute a scientist or observer.

different values of *M*. Such decoherence is robust in the sense that once a process starts the system's reduced state remains very nearly diagonal in the pointer basis throughout $R_E$. A *context of assessment* for a statement about a value of a dynamical variable *M* of a quantum system is an *M*-decoherence environment *E* of an event in which *M* may be said to take on a value. Quantum decoherence involves stabilization of a reduced quantum state, but since a quantum state is not a quantum beable (according to DP), quantum decoherence does not describe a dynamical process that *makes* a magnitude take on a particular value. The applicability of a model of environmental decoherence rather licenses one to make the meaningful statement that this magnitude takes on some value, and to find out which. Quantum measurements form an epistemically significant subset of decoherence environments. But since measurements are not the only occasions on which variables take on values, the concept of measurement need not appear in an exact formulation of quantum mechanics, according to DP. By relativizing truth to decoherence environments DP is able to secure the immanent objectivity of scientific knowledge of the quantum world. This is because all actual quantum measurements may be assessed in a single shared context.

Brukner [24] has taken recent arguments involving extensions of Wigner's friend *Gedankenexperiment* to show that in those scenarios certain facts are merely relative. For DP as well as even the modified form of RQM, such arguments present a challenge to the immanent objectivity of (relative) facts about the outcomes of quantum measurements. These relative facts either are, or are determined by, the relative truth of statements attributing dynamical properties to physical systems. A sound argument of this type proves that a relative fact in one measurement context of the scenario cannot consistently be taken to be a relative fact in some other measurement context. The contexts correspond to situations of parts of the world ("laboratories") that may be totally physically isolated from each other, but may also interact in ways modeled by unitary operations including interactions used to implement a quantum measurement on one laboratory by another. A context of assessment is the situation of one or more isolated laboratories at the conclusion of a measurement interaction within them.

DP takes such a context to be an *M*-decoherence environment for the measured variable(s) *M*. The model applies only while the laboratory in which the measurement is implemented remains isolated from external interactions: the environment then remains internal to the laboratory. A subsequent external interaction on the laboratory breaks that isolation, effectively erasing all internal measurement records. So records of all measurement outcomes are never present in all the laboratories at once, even though each laboratory at some time was assumed to contain records of the outcome of the measurement within that laboratory. A sound no-go argument then shows that not all these outcomes can be absolute facts, even though each is a relative fact in the context of the laboratory in which that outcome was obtained. According to DP [25], in the setting of the *Gedankenexperiment*, these facts are not even immanently objective.

But the *Gedankenexperiment* is not realizable in our quantum world. It requires systems involved in distinct decoherence environments to remain physically isolated from one another except during a precise, delicate action of one on another. Decoherence processes in our world are so pervasive as to prevent the required isolation. Realizing the *Gedankenexperiment* is not prevented by any physical law, but neither is its impossibility merely technological. It is ruled out by contingent but pervasive and irremovable physical features of our world. The nature of decoherence processes in our world prevents the existence of physically isolated environments like those required by the *Gedankenexperiment*. For DP, because scientists and other observers in our world share a single context of assessment, all their measurement outcomes are immanently objective facts.

**Data Availability Statement**

**Declarations**
The author did not receive support from any organization for the submitted work. The author has no relevant financial or non-financial interests to disclose.